# Multi-Longitude Observation Campaign of KV Cancri: an RR Lyrae star with irregular Blazhko modulations

Pierre de Ponthière [1]
*15 Rue Pré Mathy, Lesve – Profondeville 5170 – Belgium;*
*address email correspondence to pierredeponthiere@gmail.com*

Michel Bonnardeau
*MBCAA Observatory*
*Le Pavillon, Lalley 38930, France*

Franz-Josef (Josch) Hambsch [1,2,3]
*12 Oude Bleken, Mol, 2400 - Belgium*

Tom Krajci [1]
*P.O. Box 1351, Cloudcroft, NM 88317 - USA*

Kenneth Menzies [1]
*318A Potter Road, Framingham MA,01701 – USA*

Richard Sabo [1]
*2336 Trailcrest Drive, Bozeman MT, 59718 - USA*

1 American Association of Variable Star Observers (AAVSO)
2 Bundesdeutsche Arbeitsgemeinschaft für Veränderliche Sterne e.V. (BAV), Germany
3 Vereniging Voor Sterrenkunde (VVS), Belgium

## Abstract

We present the results of multi-longitude observations of KV Cancri, an RR Lyrae star showing an irregular Blazhko effect. With a pulsation period of 0.50208 day, the times of light curve maxima are delayed by 6 minutes per day. This daily delay regularly leads to long periods of time without maximum light curve observations for a given site. To cope with this observing time window problem, we have organized a multi-longitude observation campaign including a telescope of the AAVSONet. From the observed light curves, 92 pulsation maxima have been measured covering about six Blazhko periods. The Fourier analysis of magnitudes at maximum light has revealed a main Blazhko period of 77.6 days and also a secondary period of 40.5 days. A Fourier analysis of (O-C) values did not show the secondary Blazhko period. The frequency spectrum of the complete light curve, from a Fourier analysis and successive pre-whitening with PERIOD04, has shown triplet structures around the two Blazhko modulation frequencies but with slightly different periods (77.8 and 42.4 days). The second Blazhko frequency is statistically not a harmonic of the main Blazhko frequency. Besides the two Blazhko modulations KV Cnc presents other particularities like irregularities from Blazhko cycle to cycle and very fast magnitude variations which can reach a maximum of 2.5 magnitudes per hour over a period of 15 minutes. This campaign shows that regular observations by amateur astronomers remain important. Indeed such a detailed characterization of the Blazhko effect could not be obtained from large-scale surveys, as cooperative long time-series observations are needed.

## 1. Introduction

The designation of KV Cnc appeared in the General Catalogue of Variable Stars with the 80[th] Name List of Variable Stars (Kazarovets *et al.* 2011), and previously this star was identified as GSC 1948-1733 and NSVS 7404884. From the Northern Sky Variability Survey data (Wozniak *et al*. 2004), Wils *et al*. (2006) have measured a pulsation period of 0.50202 day and they also provided an uncertain Blazhko period of 42 days.

The current data were gathered during 158 nights between January 2012 and May 2013. During this period of 480 days, a total of 32,280 magnitude measurements covering 6 Blazhko cycles were collected. The observations were made by the authors using 20 cm to 40 cm telescopes located in Bozeman (Montana), Cloudcroft (New

Mexico), Framingham (Massachusetts), Lesve (Belgium) and Rhône-Alpes (France). The numbers of observations for the different locations are respectively 3367, 23614, 2621, 2610 and 92.

The comparison stars are given in Table 1. The star coordinates and magnitudes in B and V bands were obtained from the AAVSO's Variable Star Database (VSD). Cloudcroft observations have been reduced with C2 as a magnitude reference and C4 as a check star. The other observations have used C1 as a magnitude reference and C3 and C4 as check stars.

Since measurements were performed with V filters only, it was impossible to transform the measurements to the standard system. However, thanks to simultaneous maximum measurements from the instruments in Cloudcroft and Bozeman it has been possible to take account of the magnitude offset due to the color differences of the magnitude reference stars. This offset has been applied to observations based on C1 as reference. Dark and flat field corrections were performed with MAXIMDL software (Diffraction Limited, 2004), and aperture photometry was performed using LESVEPHOTOMETRY (de Ponthière, 2010), a custom software which also evaluates the SNR and estimates magnitude errors.

## 2. Light curve maxima analysis

The times of maxima of the light curves have been evaluated with custom software (de Ponthiere, 2010) fitting the light curve with a smoothing spline function (Reinsch, 1967). Table 2 provides the list of 92 observed maxima and Figures 1a and 1b show the (O-C) and $M_{max}$ (Magnitude at Maximum) values. From an inspection of the (O-C) and $M_{max}$ graphs the Blazhko effect is obviously irregular. The shape of the (O-C) curve during Blazhko cycles does not repeat. During the dates around HJD 245600, the (O-C) values vary more or less linearly before an abrupt fall. The $M_{max}$ graph suggests the presence of a second modulation frequency. The Blazhko effect is itself apparently modulated by a lower frequency component.

A linear regression of all available (O-C) values has provided a pulsation period of 0.5020802 d (1.99171 d$^{-1}$). The (O-C) values have been re-evaluated with this new pulsation period and the pulsation ephemeris origin has been set to the highest recorded brightness maximum: HJD 2 456 035.7518. The new derived pulsation elements are:

$$HJD_{Pulsation} = (2\ 456\ 035.7518 \pm 0.0030) + (0.5020802 \pm 0.0000078)\ E_{Pulsation}$$

The derived pulsation period is in good agreement with the value of 0.50202 published by Wils *et al.* (2006). The folded light curve on this pulsation period is shown in Figure 2.

To determine the Blazhko period, Fourier analyses and sine-wave fittings of the (O-C) values and $M_{max}$ (Magnitude at Maximum) values were performed with PERIOD04 (Lenz and Breger 2005). These analyses were limited to the first two frequency components and are given in Table 3. The frequency uncertainties have been evaluated from the Monte Carlo simulation module of PERIOD04. The obtained periods (77.02 ± 0.54 and 77.57 ± 0.25 days), for the main Blazhko effect, agree within the errors. Another secondary Blazhko period of 40.48 days is found in the spectrum of $M_{max}$ which is close to the uncertain value of 42 days provided by Wils *et al.* (2006). They probably did not detect the main Blazhko period from the Northern Sky Variability Survey due to the scarcity of the data.

On this basis the best Blazhko ephemeris is

$$HJD_{Blazhko} = 2\ 456\ 035.7518 + (77.57 \pm 0.25)\ E_{Blazhko}$$

where the origin has been selected as the epoch of the highest recorded maximum.

The (O-C) and $M_{max}$ curves folded with the Blazhko period of 77.02 days are given in Figure 3 and 4. In these diagrams, the scatter of the data is mainly due to the second Blazhko frequency and the irregular behavior from Blazhko cycle to cycle. An attempt at pre-whitening with the secondary Blazhko frequency has not significantly reduced the scatter in the folded (O-C) and $M_{max}$ graphs. Due to an irregular Blazhko effect, the two detected frequency components are not able to precisely model the (O-C) and $M_{max}$ data series.

## 3. Frequency spectrum analysis of the light curve

In the preceding paragraph, describing the $M_{max}$ analysis, a primary pulsation and two Blazhko frequencies have been identified. It will be shown that these modulating frequencies are clearly present in the spectrum of the complete light curve.

The spectrum of a signal modulated in amplitude and phase is characterized by a pattern of peaks called multiplets at the positions $kf_0 \pm nf_B$ with k and n being integers corresponding respectively to the harmonic and multiplet orders.

The frequencies, amplitudes and phases of the multiplets have been obtained with PERIOD04 by performing a succession of Fourier analyses, pre-whitenings and sine-wave fittings. Only the harmonic and multiplet components having a signal to noise ratio (SNR) greater than 4 have been retained as significant signals. Table 5 provides the complete list of Fourier components with their amplitudes, phases and uncertainties. Besides the pulsation frequency $f_0$ and harmonics $nf_0$, two series of triplets $nf_0 \pm f_B$ and $nf_0 \pm f_{B2}$ based on the principal and secondary Blazhko frequencies $f_B$ and $f_{B2}$ have been found. The Blazhko frequencies and corresponding periods are tabulated in Table 4 with their uncertainties. These Blazhko periods are close to the values obtained with $M_{Max}$ analysis given in Table 3. During the sine-wave fitting, the fundamental frequency $f_0$ and largest triplets $f_0 + f_B$ and $f_0 + f_{B2}$ have been left unconstrained and the other frequencies have been entered as combinations of these three frequencies. The uncertainties of frequencies, amplitudes and phases have been estimated by Monte Carlo simulations. The amplitude and phase uncertainties have been multiplied by a factor of two as it is known that the Monte Carlo simulations underestimate these uncertainties (Kolenberg et al. 2009). The two Blazhko modulation frequencies $f_B$ (0.012853) and $f_{B2}$ (0.023593) are statistically ($\sigma$ = 16x10$^{-6}$) not in resonance, provided that the n:m resonance ratios with n or m greater than 10 are not taken into account. For CZ Lacertae, Sódor et al.(2011) and V784 Ophiuchi, de Ponthière et al. (2013) 5:4 and 5:6 resonance ratios have been found.

Table 6 lists for each harmonic the amplitude ratios $A_i/A_1$ and the ratios usually used to characterize the Blazhko effect, that is, $A_i^+/A_1$ ; $A_i^-/A_1$ ; $R_i = A_i^+/A_i^-$ and asymmetries $Q_i = (A_i^+ - A_i^-)/(A_i^+ + A_i^-)$. In the present case the side lobe $A_1^-$ is larger than $A_1^+$ which leads to a negative value (-0.13) for the $Q_1$ asymmetry ratio. It is not unusual but for the majority of the Blazhko stars, this asymmetry ratio is positive (see figure 10 of Alcock et al. 2003). The $R_i$ and $Q_i$ ratios for triplets around the secondary Blazhko frequency $f_{B2}$ are also given in Table 6. The asymmetry ratios $Q_i$ for $f_{B2}$ are positive.

## 4. Light curve variations over Blazhko cycle

Subdividing the data set into temporal subsets is a classical method to visualize and analyze the light curve variations over the Blazhko cycle. Ten temporal subsets corresponding to the different Blazhko phase intervals $\Psi_i$ (i = 0 , 9) have been created using the epoch of the highest recorded maximum (2456035.7518) as the origin of the first subset. The folded light curves for the ten subsets are presented in Figure 5. Over the subsets, the number of data points varies between 1916 and 5678. Other than a lack of coverage in two subsets when the light curve is at its minimum, the data points are relatively well distributed.

Despite the subdivision over the Blazhko cycle, a scatter still remains on the light curves; this fact has been already pointed out in the light curve maxima analysis. A visual inspection of the light curves in different subsets reveals that the light curve slope is at its steepest value in the two subsets from Blazhko phases 0.9 to 0.1, that is, when the peak to peak magnitude variations and magnitude at maximum are at their maximal values. An astonishing slope of 2.9 magnitudes per hour has been recorded. Generally the RR Lyrae light curves present a bump just before the minimum. For KV Cnc, in the two subsets from 0.9 to 0.1, the bump is replaced by a slightly increasing slope.

Fourier analyses and Least-Square fittings have been performed on the different temporal subsets. For the fundamental and the first four harmonics the amplitudes $A_i$ and the epoch-independent phase differences ($\Phi_{k1} = \Phi_k - k\Phi_{k1}$) are given in Table 7 and plotted in Figure 6. The amplitudes have large uncertainties for the subsets 0.2 - 0.3 and 0.3 - 0.4. This is due to the lack of coverage at light curve minimum as shown in Figure 5. These amplitude uncertainties probably impact the epoch-independent phase differences especially in the subset 0.3 – 0.4 where the phase differences seem to be dubious. As expected the $A_1$ amplitudes of the fundamental frequency have lower values for Blazhko phases 0.4 to 0.7, that is, when the light curve amplitude variations on the pulsation are weaker.

## 5. Conclusions

Blazhko modulations have been detected by measurements of (O-C) values and magnitude of light curve maxima and confirmed by complete light curve Fourier analysis. The Blazhko periods obtained by the complete light curve analysis are reported as their period uncertainties are lower. The main Blazhko period (1 / $f_B$) is 77.80 ± 0.05 days. The secondary Blazhko period (1 / $f_{B2}$) is 42.39 ± 0.03 days. These two Blazhko modulations are not in resonance. Regular and coordinated multi-longitude observations by amateurs have been needed to cope with the problem of observation time windows created by the pulsation period of 0.50208 days. Amateur astronomers observing RR Lyrae stars have the tendency to restrict their observations near the maximum of light curve which is indeed important. However, the problems encountered in the Fourier analysis in Blazhko subsets were due to a lack of data during the minimum part of the pulsation cycle. Observers are encouraged to also image during pulsation phases other than near the maximum.

## Acknowledgements

Dr A. Henden, Director of AAVSO and the AAVSO are acknowledged for the use of AAVSOnet telescopes at Cloudcroft (New Mexico, USA). The authors thank the referee for constructive comments which have helped to clarify and improve the paper. This work has made use of The International Variable Star Index (VSX) maintained by the AAVSO and of the SIMBAD astronomical database (http://simbad.u-strasbg.fr).

## Table 1. Comparison stars

| Identification | AAVSO AUID | R.A. (2000) h m s | Dec (2000) ° ' " | B | V | B-V | |
|---|---|---|---|---|---|---|---|
| GSC 1948-1556 | 000-BKL-413 | 08 40 05.47 | +27 39 12.1 | 12.526 | 11.998 | 0.528 | C1 |
| GSC 1948-1451 | 000-BKL-415 | 08 40 09.30 | +27 41 19.4 | 13.627 | 12.946 | 0.681 | C2 |
| GSC 1948-1631 | 000-BKL-416 | 08 40 34.19 | +27 47 50.0 | 13.972 | 13.204 | 0.768 | C3 |
| GSC 1948-1548 | 000-BKL-417 | 08 40 00.87 | +27 42 35.9 | 14.110 | 13.543 | 0.567 | C4 |

## Table 2. List of measured maxima

| Maximum HJD | Error | O-C (day) | E | Magnitude (V) | Error | Location |
|---|---|---|---|---|---|---|
| 2455943.4189 | 0.0026 | 0.0553 | -184 | 12.023 | 0.01 | 2 |
| 2455944.4178 | 0.0030 | 0.0500 | -182 | 12.007 | 0.01 | 2 |
| 2455960.4523 | 0.0012 | 0.0170 | -150 | 11.663 | 0.01 | 2 |
| 2455962.4642 | 0.0018 | 0.0205 | -146 | 11.709 | 0.012 | 2 |
| 2455967.5064 | 0.0021 | 0.0416 | -136 | 11.930 | 0.003 | 3 |
| 2455970.5282 | 0.0026 | 0.0507 | -130 | 11.990 | 0.005 | 3 |
| 2455971.5335 | 0.0035 | 0.0518 | -128 | 11.990 | 0.004 | 3 |
| 2455978.5639 | 0.0016 | 0.0526 | -114 | 12.144 | 0.004 | 3 |
| 2455984.5914 | 0.0026 | 0.0548 | -102 | 12.220 | 0.004 | 3 |
| 2455989.6193 | 0.0053 | 0.0616 | -92 | 12.278 | 0.005 | 4 |
| 2455992.6310 | 0.0037 | 0.0607 | -86 | 12.276 | 0.006 | 3 |
| 2455996.6553 | 0.0038 | 0.0681 | -78 | 12.288 | 0.004 | 5 |
| 2455997.6525 | 0.0039 | 0.0611 | -76 | 12.249 | 0.005 | 3 |
| 2455998.6595 | 0.0047 | 0.0638 | -74 | 12.285 | 0.005 | 4 |
| 2456000.6733 | 0.0077 | 0.0692 | -70 | 12.255 | 0.005 | 4 |
| 2456004.6854 | 0.0043 | 0.0644 | -62 | 12.255 | 0.005 | 3 |
| 2456008.7018 | 0.0034 | 0.0639 | -54 | 12.245 | 0.005 | 3 |
| 2456008.7032 | 0.0029 | 0.0653 | -54 | 12.243 | 0.004 | 5 |
| 2456008.7041 | 0.0043 | 0.0662 | -54 | 12.243 | 0.005 | 4 |
| 2456009.7026 | 0.0078 | 0.0605 | -52 | 12.277 | 0.005 | 4 |
| 2456009.7085 | 0.0040 | 0.0664 | -52 | 12.277 | 0.004 | 5 |
| 2456011.7174 | 0.0065 | 0.0669 | -48 | 12.245 | 0.005 | 4 |
| 2456011.7219 | 0.0035 | 0.0714 | -48 | 12.243 | 0.004 | 5 |
| 2456013.7404 | 0.0043 | 0.0814 | -44 | 12.250 | 0.005 | 4 |
| 2456015.7505 | 0.0039 | 0.0831 | -40 | 12.204 | 0.005 | 4 |
| 2456016.7472 | 0.0030 | 0.0756 | -38 | 12.185 | 0.006 | 4 |
| 2456021.7383 | 0.0017 | 0.0456 | -28 | 11.999 | 0.005 | 5 |
| 2456023.7350 | 0.0018 | 0.0338 | -24 | 11.947 | 0.007 | 4 |
| 2456028.7337 | 0.0007 | 0.0114 | -14 | 11.703 | 0.004 | 5 |
| 2456029.7368 | 0.0013 | 0.0103 | -12 | 11.696 | 0.005 | 4 |
| 2456030.7368 | 0.0010 | 0.0061 | -10 | 11.617 | 0.019 | 5 |
| 2456031.7412 | 0.0018 | 0.0063 | -8 | 11.590 | 0.039 | 5 |
| 2456033.7459 | 0.0007 | 0.0025 | -4 | 11.541 | 0.006 | 4 |
| 2456034.7501 | 0.0020 | 0.0025 | -2 | 11.545 | 0.006 | 4 |
| 2456035.7518 | 0.0030 | 0.0000 | 0 | 11.524 | 0.005 | 4 |
| 2456038.7683 | 0.0006 | 0.0038 | 6 | 11.614 | 0.006 | 4 |
| 2456039.7718 | 0.0005 | 0.0031 | 8 | 11.622 | 0.004 | 5 |
| 2456039.7740 | 0.0006 | 0.0053 | 8 | 11.615 | 0.006 | 4 |
| 2456055.3917 | 0.0026 | 0.0576 | 39 | 12.106 | 0.012 | 2 |
| 2456061.4163 | 0.0052 | 0.0569 | 51 | 12.165 | 0.016 | 2 |
| 2456202.9511 | 0.0012 | -0.0033 | 333 | 11.862 | 0.006 | 4 |
| 2456205.9689 | 0.0011 | 0.0018 | 339 | 11.969 | 0.005 | 4 |
| 2456209.0013 | 0.0040 | 0.0215 | 345 | 12.051 | 0.011 | 5 |
| 2456248.6817 | 0.0039 | 0.0353 | 424 | 12.048 | 0.011 | 2 |
| 2456254.6918 | 0.0015 | 0.0200 | 436 | 11.861 | 0.005 | 3 |
| 2456256.6971 | 0.0018 | 0.0169 | 440 | 11.815 | 0.004 | 3 |
| 2456265.7291 | 0.0013 | 0.0109 | 458 | 11.777 | 0.004 | 3 |
| 2456268.7406 | 0.0016 | 0.0098 | 464 | 11.777 | 0.004 | 3 |
| 2456276.7683 | 0.0012 | 0.0037 | 480 | 11.724 | 0.016 | 3 |
| 2456279.7819 | 0.0015 | 0.0046 | 486 | 11.730 | 0.013 | 4 |
| 2456280.7878 | 0.0004 | 0.0063 | 488 | 11.767 | 0.004 | 5 |
| 2456281.7901 | 0.0017 | 0.0044 | 490 | 11.740 | 0.014 | 4 |
| 2456282.7959 | 0.0013 | 0.0060 | 492 | 11.778 | 0.01 | 4 |
| 2456287.8229 | 0.0015 | 0.0119 | 502 | 11.924 | 0.004 | 3 |
| 2456290.8423 | 0.0023 | 0.0186 | 508 | 11.960 | 0.014 | 4 |
| 2456292.8584 | 0.0031 | 0.0263 | 512 | 12.062 | 0.006 | 3 |
| 2456294.8715 | 0.0026 | 0.0309 | 516 | 12.109 | 0.005 | 3 |
| 2456295.8814 | 0.0023 | 0.0366 | 518 | 12.106 | 0.005 | 3 |
| 2456297.9004 | 0.0045 | 0.0472 | 522 | 12.158 | 0.007 | 5 |
| 2456300.9255 | 0.0029 | 0.0596 | 528 | 12.202 | 0.007 | 3 |
| 2456303.4492 | 0.0041 | 0.0728 | 533 | 12.168 | 0.011 | 2 |
| 2456308.9646 | 0.0043 | 0.0650 | 544 | 12.125 | 0.019 | 4 |
| 2456308.9684 | 0.0096 | 0.0688 | 544 | 12.146 | 0.073 | 5 |
| 2456310.9718 | 0.0034 | 0.0637 | 548 | 12.106 | 0.01 | 4 |
| 2456311.9694 | 0.0030 | 0.0571 | 550 | 12.117 | 0.011 | 5 |
| 2456311.9742 | 0.0042 | 0.0619 | 550 | 12.101 | 0.017 | 4 |
| 2456312.9772 | 0.0068 | 0.0607 | 552 | 12.079 | 0.013 | 4 |
| 2456313.9725 | 0.0027 | 0.0518 | 554 | 12.064 | 0.011 | 4 |
| 2456314.9747 | 0.0028 | 0.0497 | 556 | 12.077 | 0.011 | 5 |
| 2456315.9759 | 0.0023 | 0.0467 | 558 | 12.028 | 0.011 | 4 |
| 2456323.4865 | 0.0041 | 0.0257 | 573 | 11.993 | 0.02 | 2 |
| 2456323.9876 | 0.0035 | 0.0247 | 574 | 11.980 | 0.017 | 4 |
| 2456328.4990 | 0.0028 | 0.0171 | 583 | 12.007 | 0.008 | 3 |
| 2456330.5028 | 0.0016 | 0.0124 | 587 | 11.975 | 0.004 | 3 |
| 2456334.5124 | 0.0017 | 0.0051 | 595 | 11.937 | 0.004 | 3 |
| 2456340.5428 | 0.0015 | 0.0102 | 607 | 12.000 | 0.008 | 1 |
| 2456343.5518 | 0.0015 | 0.0066 | 613 | 11.984 | 0.009 | 1 |
| 2456351.5974 | 0.0035 | 0.0184 | 629 | 11.734 | 0.019 | 4 |
| 2456352.6018 | 0.0036 | 0.0186 | 631 | 11.726 | 0.022 | 4 |
| 2456353.6024 | 0.0021 | 0.0150 | 633 | 11.682 | 0.023 | 4 |
| 2456354.6071 | 0.0020 | 0.0154 | 635 | 11.679 | 0.016 | 4 |
| 2456358.6164 | 0.0019 | 0.0079 | 643 | 11.645 | 0.016 | 4 |
| 2456363.6383 | 0.0018 | 0.0087 | 653 | 11.825 | 0.011 | 5 |
| 2456363.6391 | 0.0025 | 0.0095 | 653 | 11.784 | 0.024 | 4 |
| 2456369.6638 | 0.0015 | 0.0088 | 665 | 12.022 | 0.004 | 3 |
| 2456374.7017 | 0.0026 | 0.0256 | 675 | 12.188 | 0.006 | 3 |
| 2456376.7208 | 0.0073 | 0.0363 | 679 | 12.257 | 0.026 | 5 |
| 2456392.8007 | 0.0072 | 0.0487 | 711 | 12.048 | 0.014 | 5 |
| 2456400.8163 | 0.0015 | 0.0305 | 727 | 11.944 | 0.007 | 5 |
| 2456403.3272 | 0.0024 | 0.0309 | 732 | 11.868 | 0.012 | 2 |
| 2456407.3363 | 0.0025 | 0.0231 | 740 | 11.885 | 0.01 | 2 |
| 2456418.3574 | 0.0037 | -0.0022 | 762 | 11.970 | 0.012 | 2 |

*Locations : 1 – Rhône-Alpes (France); 2 – Lesve (Belgium); 3 – Framingham (MA); 4 – Cloudcroft (NM); 5 –Bozeman (MT)*

**Table 3. Blazhko spectral components from light curve maxima**

*From (O-C) values*

| Frequency (cycle/days) | σ($d^{-1}$) | Period (days) | σ(d) | Amplitude (days) | Φ (cycle) | SNR |
|---|---|---|---|---|---|---|
| 0.01298 | 9 $10^{-5}$ | 77.02 | 0.54 | 0.028 | 0.926 | 10.4 |

*From $M_{max}$*

| Frequency (cycle/days) | σ($d^{-1}$) | Period (days) | σ(d) | Amplitude (V mag) | Φ (cycle) | SNR |
|---|---|---|---|---|---|---|
| 0.01289 | 4 $10^{-5}$ | 77.57 | 0.25 | 0.245 | 0.993 | 22.6 |
| 0.02471 | 9 $10^{-5}$ | 40.48 | 0.14 | 0.119 | 0.398 | 11.0 |

**Table 4 Blazhko frequencies and periods derived from triplets**

| Component | Derived from | Frequency ($d^{-1}$) | σ($d^{-1}$) | Period (d) | σ(d) |
|---|---|---|---|---|---|
| $f_0$ | | 1.991689 | 1.7x$10^{-6}$ | 0.5020864 | 4x$10^{-7}$ |
| $f_B$ | $f_0 + f_B$ | 0.012853 | 8x$10^{-6}$ | 77.80 | 0.05 |
| $f_{B2}$ | $f_0 + f_{B2}$ | 0.023593 | 16x$10^{-6}$ | 42.39 | 0.03 |

## Table 5 Multi-frequency fit results

| Component | $f(d^{-1})$ | $\sigma(f)$ | $A_i$ (mag) | $\sigma(A_i)$ | $\Phi_i$ (cycle) | $\sigma(\Phi_i)$ | SNR |
|---|---|---|---|---|---|---|---|
| $f_0$ | 1.991689 | $1.7 \times 10^{-6}$ | 0.3646 | 0.0008 | 0.2022 | 0.0005 | 118.5 |
| $2 f_0$ | 3.983378 | | 0.1298 | 0.0012 | 0.7825 | 0.0013 | 46.0 |
| $3 f_0$ | 5.975067 | | 0.0818 | 0.0011 | 0.4018 | 0.0018 | 29.8 |
| $4 f_0$ | 7.966757 | | 0.0400 | 0.0009 | 0.0400 | 0.0046 | 16.4 |
| $5 f_0$ | 9.958446 | | 0.0268 | 0.0010 | 0.6656 | 0.0065 | 13.5 |
| $6 f_0$ | 11.950135 | | 0.0203 | 0.0011 | 0.3226 | 0.0082 | 11.7 |
| $7 f_0$ | 13.941824 | | 0.0110 | 0.0009 | 0.0354 | 0.0119 | 7.0 |
| $8 f_0$ | 15.933513 | | 0.0055 | 0.0008 | 0.7105 | 0.0230 | 3.9 |
| $f_0 + f_B$ | 2.004542 | $8 \times 10^{-6}$ | 0.0582 | 0.0011 | 0.5060 | 0.0031 | 18.9 |
| $f_0 - f_B$ | 1.978836 | | 0.0759 | 0.0009 | 0.0580 | 0.0017 | 24.7 |
| $2 f_0 + f_B$ | 3.996231 | | 0.0482 | 0.0010 | 0.1499 | 0.0040 | 17.1 |
| $2 f_0 - f_B$ | 3.970525 | | 0.0419 | 0.0011 | 0.6413 | 0.0044 | 14.8 |
| $3 f_0 + f_B$ | 5.987920 | | 0.0303 | 0.0011 | 0.7451 | 0.0060 | 11.0 |
| $3 f_0 - f_B$ | 5.962215 | | 0.0214 | 0.0010 | 0.2807 | 0.0076 | 7.8 |
| $4 f_0 + f_B$ | 7.979609 | | 0.0216 | 0.0011 | 0.3640 | 0.0076 | 8.9 |
| $4 f_0 - f_B$ | 7.953904 | | 0.0244 | 0.0011 | 0.8420 | 0.0071 | 9.9 |
| $5 f_0 + f_B$ | 9.971299 | | 0.0130 | 0.0009 | 0.9890 | 0.0116 | 6.6 |
| $5 f_0 - f_B$ | 9.945593 | | 0.0211 | 0.0011 | 0.5055 | 0.0079 | 10.7 |
| $6 f_0 + f_B$ | 11.962988 | | 0.0096 | 0.0008 | 0.5410 | 0.0144 | 5.5 |
| $6 f_0 - f_B$ | 11.937282 | | 0.0150 | 0.0011 | 0.1600 | 0.0119 | 8.7 |
| $7 f_0 + f_B$ | 13.954677 | | 0.0096 | 0.0009 | 0.1890 | 0.0115 | 6.1 |
| $7 f_0 - f_B$ | 13.92897 | | 0.0089 | 0.0007 | 0.8373 | 0.0154 | 5.7 |
| $f_0 + f_{B2}$ | 2.015282 | $16 \times 10^{-6}$ | 0.0394 | 0.0008 | 0.2681 | 0.0034 | 12.8 |
| $f_0 - f_{B2}$ | 1.968096 | | 0.0258 | 0.0009 | 0.7453 | 0.0066 | 8.4 |
| $2 f_0 + f_{B2}$ | 4.006971 | | 0.0352 | 0.0009 | 0.8826 | 0.0042 | 12.5 |
| $2 f_0 - f_{B2}$ | 3.959785 | | 0.0124 | 0.0010 | 0.1666 | 0.0113 | 4.3 |
| $3 f_0 + f_{B2}$ | 5.998660 | | 0.0174 | 0.0010 | 0.5395 | 0.0084 | 6.4 |
| $3 f_0 - f_{B2}$ | 5.951475 | | 0.0154 | 0.0009 | 0.7617 | 0.0102 | 5.6 |

## Table 6. KV Cnc Harmonic, Triplet amplitudes, ratios and asymmetry parameters

| i | $A_i/A_1$ | $A_i^+/A_1$ | $A_i^-/A_1$ | $R_i$ | $Q_i$ | $R_i (f_{B2})$ | $Q_i (f_{B2})$ |
|---|---|---|---|---|---|---|---|
| 1 | 1.00 | 0.16 | 0.21 | 0.77 | -0.13 | 1.53 | 0.21 |
| 2 | 0.36 | 0.13 | 0.11 | 1.15 | 0.07 | 2.85 | 0.48 |
| 3 | 0.22 | 0.08 | 0.06 | 1.42 | 0.17 | 1.13 | 0.06 |
| 4 | 0.11 | 0.06 | 0.07 | 0.89 | -0.06 | - | - |
| 5 | 0.07 | 0.04 | 0.06 | 0.62 | -0.24 | - | - |
| 6 | 0.06 | 0.03 | 0.04 | 0.64 | -0.22 | - | - |
| 7 | 0.03 | 0.03 | 0.02 | 1.08 | 0.04 | - | - |
| 8 | 0.02 | - | - | - | - | - | - |

## Table 7. KV Cnc Fourier coefficients over Blazhko cycle
### based on period of 77.80 days

| $\Psi$ (cycle) | $A_1$ (mag) | $A_2$ (mag) | $A_3$ (mag) | $A_4$ (mag) | $\Phi_1$ (rad) | $\Phi_{21}$ (rad) | $\Phi_{31}$ (rad) | $\Phi_{41}$ (rad) |
|---|---|---|---|---|---|---|---|---|
| 0.0 - 0.1 | 0.503 | 0.190 | 0.117 | 0.082 | 1.410 | 2.315 | 4.860 | 1.300 |
| 0.1 - 0.2 | 0.467 | 0.187 | 0.133 | 0.082 | 1.356 | 2.430 | 5.207 | 1.953 |
| 0.2 - 0.3 | 0.420 | 0.090 | 0.096 | 0.076 | 1.142 | 2.487 | 4.463 | 1.278 |
| 0.3 - 0.4 | 0.215 | 0.201 | 0.068 | 0.093 | 0.398 | 4.022 | 4.811 | 3.762 |
| 0.4 - 0.5 | 0.244 | 0.099 | 0.036 | 0.021 | 1.206 | 2.213 | 5.651 | 1.804 |
| 0.5 - 0.6 | 0.296 | 0.133 | 0.068 | 0.033 | 1.148 | 2.308 | 5.326 | 1.332 |
| 0.6 - 0.7 | 0.279 | 0.134 | 0.080 | 0.039 | 0.996 | 2.315 | 5.350 | 1.856 |
| 0.7 - 0.8 | 0.345 | 0.156 | 0.094 | 0.044 | 0.975 | 2.216 | 5.103 | 1.596 |
| 0.8 - 0.9 | 0.402 | 0.141 | 0.089 | 0.048 | 1.309 | 2.230 | 4.587 | 0.987 |
| 0.9 - 1.0 | 0.482 | 0.191 | 0.115 | 0.075 | 1.409 | 2.405 | 5.124 | 1.666 |

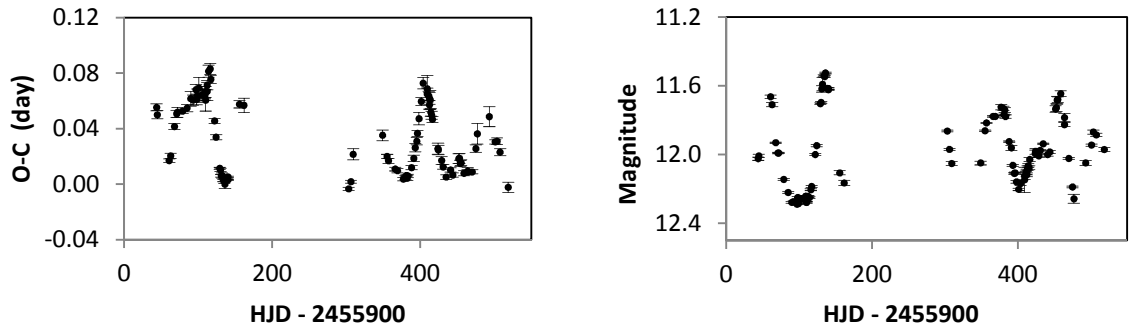

Figure 1. KV Cnc O-C (days) and Magnitude at maximum

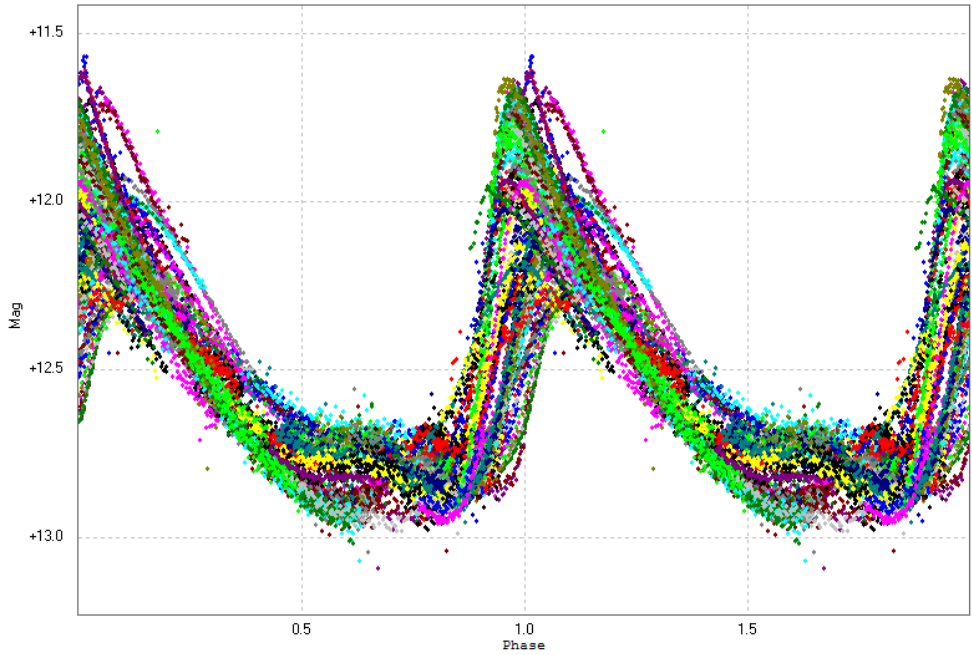

Figure 2. KV Cnc light curve folded with pulsation period

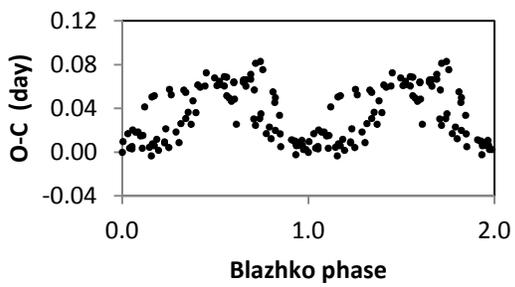

Figure 3. KV Cnc O-C
based on Blazhko period of 77.57 days

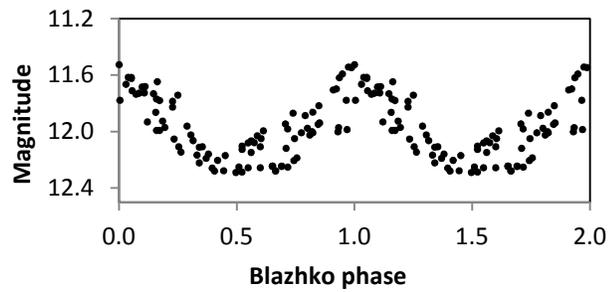

Figure 4. KV Cnc Magnitude at maximum
based on Blazhko period of 77.57 days

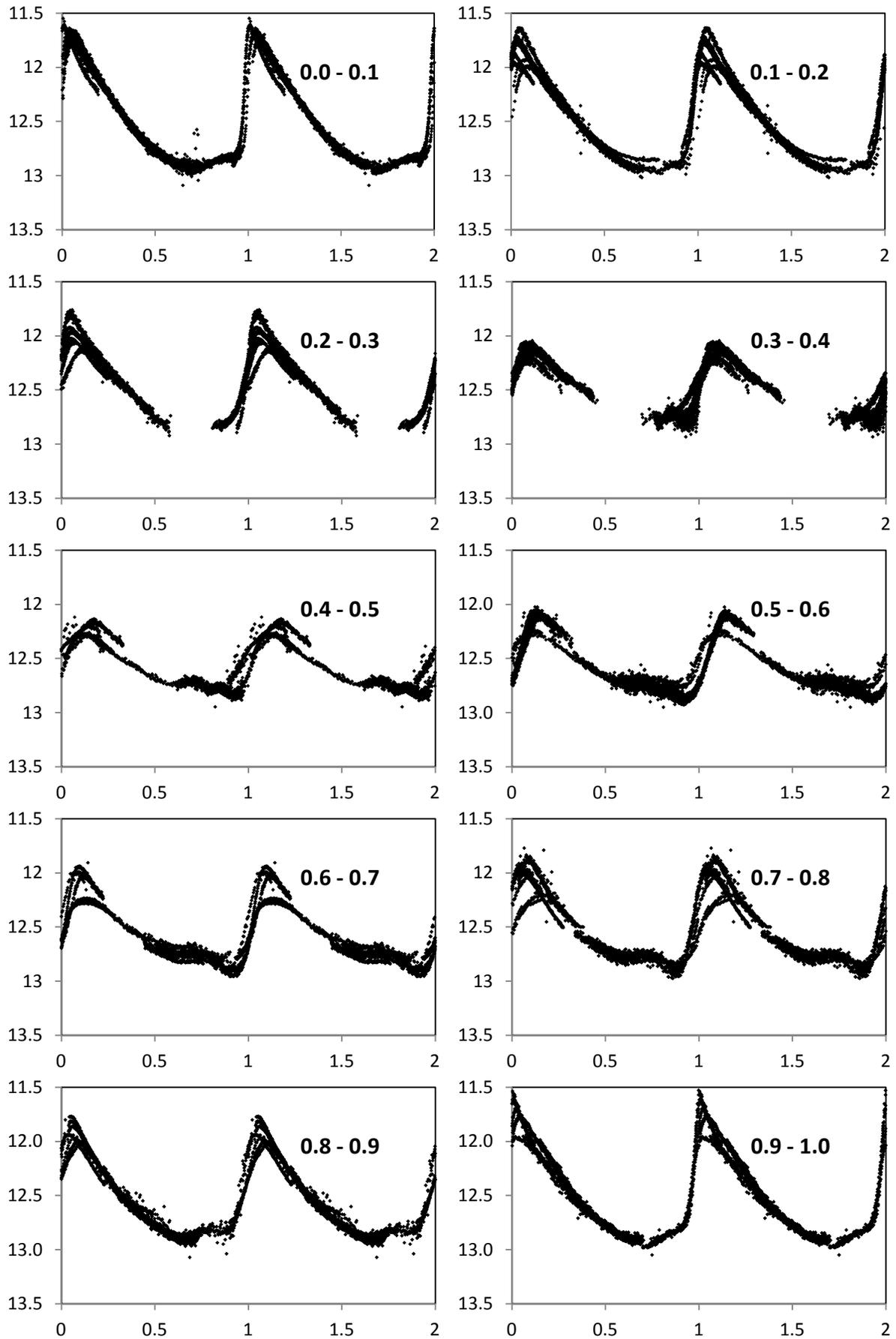

Figure 5 KV Cnc light curves for different temporal subsets (magnitude vs. pulsation phase) based on a Blazhko period of 77.80 days.

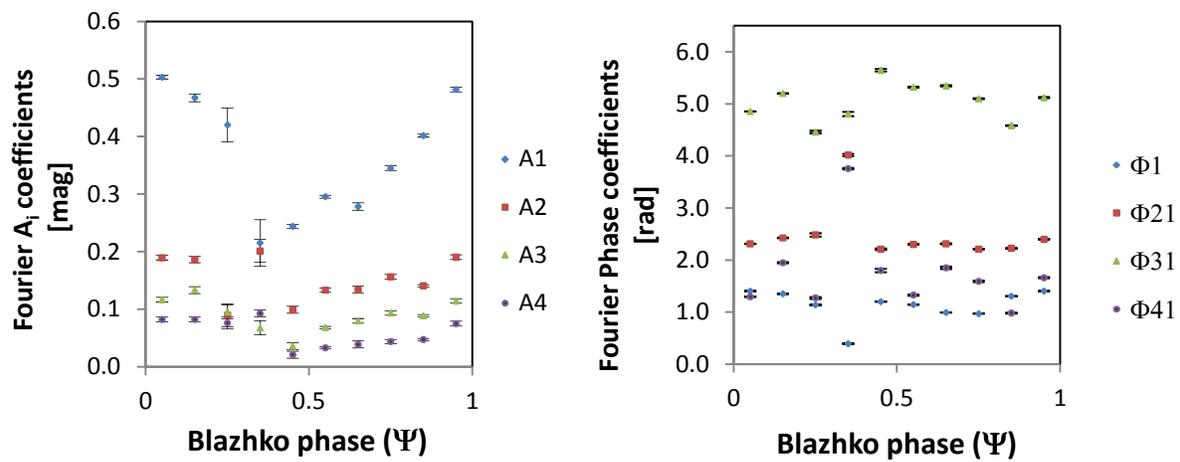

Figure 6a (left). KV Cnc Fourier $A_i$ amplitude (mag) for the ten temporal subsets based on a Blazhko period of 77.80 days.
Figure 6b (right) Fourier $\Phi_1$ and $\Phi_{ki}$ phase (rad) for the ten temporal subsets based on a Blazhko period of 77.80 days.